# Using statistical techniques and replication samples for imputation of metabolite missing values

Akram Yazdani[1*] and Azam Yazdani[2]

[1] Department of Genetics and Genomic Sciences, Icahn School of Medicine at Mount Sinai, New York, 10029, USA

[2] Human Genetics Center, Department of Epidemiology, 1200 Pressler Street, Suite E-523, Houston, Texas, 77030, USA

Address correspondence to:

Akram Yazdani, PhD

akramyazdani16@gmail.com

## Abstract

**Background**: Data preparation, such as missing values imputation and transformation, is the first step in any data analysis and requires crucial attention. Particularly, analysis of metabolites demands more preparation since those small compounds have recently been measurable in large scales with mass spectrometry techniques. We introduce novel statistical techniques for metabolite missing values imputation by utilizing replication samples.

**Results**: To understand the nature of the missing values using replication samples, we obtained the empirical distribution of missing values and observed that the rate of missing values is approximately distributed as uniform across the metabolite range. Therefore, the missing values cannot be imputed with the lowest values. Using the identified distribution, we illustrated a simulation study to find an optimal imputation approach for metabolites.

**Conclusions**: We demonstrated that the missing values in metabolomic data sets might not be necessarily low value. After identification of the nature of missing values, we validated K nearest neighborhood as an optimal approach for imputation.

**Key words:** Statistical techniques, missing value imputation, empirical distribution, optimal imputation, metabolomic

## Background

Extracting relevant and substantial biological information from large-scale datasets at different biological levels is one of the challenges in modern biomedical research. The extraction process is highly dependent on data preparation, including imputation and data transformation to hold underlying assumptions. To provide a valid analysis, missing value imputation techniques need to be carefully selected since using inappropriate imputation approaches results in a loss of power and misleading conclusions. Therefore, the nature of missing values and an optimal imputation approach need to be identified before initiating any data analysis.

Missing values widely occur in mass spectrometry metabolomics datasets due to a variety of reasons, such as values that exist below the detection limit of the mass spectrometer or technical issues unrelated to the metabolite processing [1-3]. The first step of preprocessing raw metabolomic data includes baseline correction, noise reduction, smoothing, peak detection, and alignment [4-7]. After the preprocessing and applying stringent quality control standards for metabolite levels by the laboratory, some other steps such as imputation and transformation are required to prepare the data for analysis.

Missing value imputation has been addressed through a wide range of approaches, including disregarding all variables with missing values or using univariate, multivariate, Bayesian approaches [e.g. 1, 6, 8-9]. Some studies highlight the importance of imputation of missing values in data processing pipelines by demonstrating the major effect of missing data estimation algorithms on the outcome of data analysis [e.g. 4, 10-11]. However, imputing metabolomic missing values with the lowest value of measured metabolites is going to be a common approach. Therefore, new techniques are required for the identification of the nature of missing values to emphasize the importance of the choice of imputation approach.

To identify the nature of missing values, we take advantage of the availability of replication samples that occurs in many biomedical and biological studies. Using replication samples, we proposed new statistical techniques to obtain the empirical distribution of missing data. Through the application of the techniques on 97 replication samples in metabolomics data, we identified the empirical distribution of missing values and showed the missing values are not under the detection limit. We also conducted a simulation study based on the identified distribution of missing values for 1977 individuals and found that the KNN algorithm performs better than the other approaches. The result of this study provides validation to other studies that made the same conclusion through different approaches.

## Methods and Results

**Study sample**: Our metabolomic data were collected on a subset of the Atherosclerosis Risk in Communities (ARIC) study (The ARIC Investigators 1989) [12]. Metabolite profiling was completed in June 2010 using fasting serum samples that had been stored at -80°C since collection at the baseline examination in 1987–1989. For the discovery of African-American samples, detection and quantification of metabolites were completed, using an untargeted, gas chromatography-mass spectrometry, GC-MS, and liquid chromatography-mass spectrometry, LC-MS, based metabolomic quantification protocol [13-14]. Pre-processing of the raw data, including baseline correction, noise reduction, smoothing, peak detection alignment, was carried out by Metabolon Inc., and 608 metabolites were recorded over 1977 individuals. Since the source of missing metabolomic data varies from biological to technical reasons, using statistical techniques here, we focus on the imputation of metabolites that have 50 percent or less missing values.

Based on the requirement for each stage of analysis, we used a particular subset of data with specific features. For instance, we used replication samples of 97 individuals who shared 159 metabolites to identify the nature of missing values with a novel approach. Since these replication samples were acquired 4-5 years apart, we first determined the effect of 4-5 years longer freezing on serum metabolites. We note here that the metabolites were measured from frozen serum more than 20 years, which may have already resulted in the loss of several metabolites. However, we wanted to address whether 4-5 years longer freezing (after 20 years total freezing) has a significant impact on metabolite loss.

**Effect of freezing:** By comparing the empirical distributions of 39 metabolites with no missing values in replication set, we evaluated whether the effect of longer freezing up to 5 years was significant. By comparing the empirical distributions of 39 metabolites with no missing values in replication set, we evaluated whether the effect of longer freezing up to 5 years was significant. To see if the measured metabolites in two different time points were comparable at an individual level, we plotted the replication samples at one time point versus the other time point for each metabolite, Supplementary1. The plots do not represent any significant evidence of differences or trends associated with the time points, although they might be slightly different. We employed parametric approaches for the set of metabolites that were normally distributed and nonparametric approaches for metabolites that were not normal. For the set of metabolites with a normal distribution, t-test and F-test were applied to assess whether the distribution of a metabolite in replication samples was statistically the same. For the set of non-normal metabolites, we applied the two-sample Kolmogorov-Smirnov Test. Using these approaches, the null hypothesis was not rejected at level 0.05, which means there is no significant difference between metabolites that are frozen for ~20 years and those that are frozen for 4-5 years longer (~25 years). Furthermore, we calculated the Kullback-Leibler divergence (KLD) presented for each metabolite in Supplementary1 that ranges from 0.02 to 0.271. The calculated KLDs close to zero reveal the similar distribution of replication samples at both time points which are inconsistency with the other results and plots.

**Empirical distribution of missing values:** Since the missing values in each metabolite did not provide enough sample for making any conclusion, we pooled all the missing values that are observed through replication to assess how they were distributed across the range of metabolites. In order to provide sufficient conditions that allow us to pool missing values, we carried out some assessments.

The first assessment was related to the distribution of metabolites. We noticed that different transformations were required to transform metabolites to normal distribution. Therefore, we selected 47 metabolites that were normally distributed using the same transformation. We then standardized the metabolites in order to pool their missing values. The second assessment was related to the missing values that were not observed in any replication. Table1 shows the number of those missing values. We excluded those missing values from the analysis at this step and reached to a set of 575 missing values imputed with replication samples.

**Table 1.** Number of missing values remained non-observed in replication samples.

| Metabolite | Trehalose | Theophylline | Stearoylcarnitine | Phenylacetate | Glycodeoxycholate |
|---|---|---|---|---|---|
| Non-observed missing values | 2 | 4 | 1 | 3 | 4 |

We pooled missing values across 47 metabolites normally distributed with the same transformation and estimated the distribution of missing values using Kolmogorov-Smirnov Goodness-of-fit. The *p*-value 0.026 reveals that the missing values are approximately normally distributed, Figure 1. While the metabolites had been transformed to the normal distribution, we calculated the rate of missing values in the same intervals obtained based on quartile of missing values. We observed that the rate of missing values was approximately uniform over the range of metabolite values, Figure 2. Although, the rate of missing values in the first quartile is slightly higher than the others.

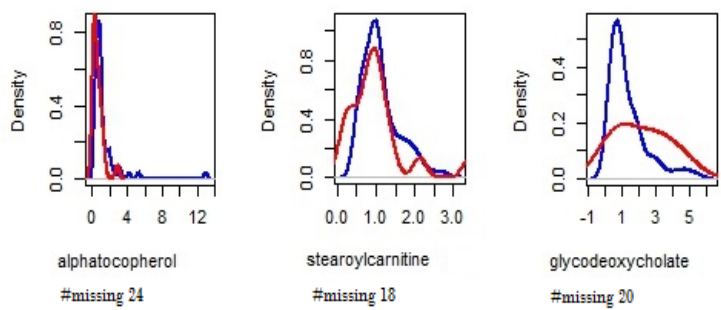

**Figure 1.** Distribution of missing metabolites observed through replication. The null hypothesis of the normal distribution of missing values is not rejected using Kolmogorov-Smirnov Goodness-of-fit at level 0.01.

The assumption that metabolites are missing because they are low is not supported by this analysis. Therefore, replacing missing values with the lowest values can severely distort the distribution of metabolites and result in misleading and inaccurate conclusions.

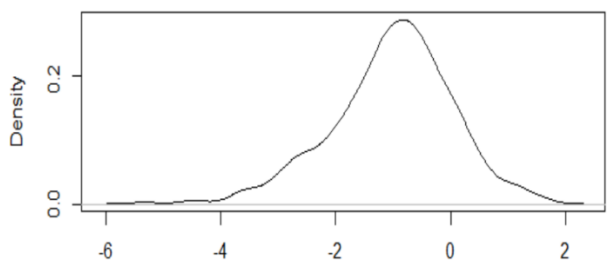

**Figure 2.** Rate of missing values across the range of metabolites. $Qi$ in x-axis shows $i$th quartile of missing values. Y-axis shows the rate of missing values in each interval.

We also plotted the empirical distribution of each metabolite for missing and observed values separately. Figure 3 demonstrates the distributions for three metabolites, while the distributions of the entire set are provided in Supplementary2. The distributions showed that the missing values were distributed across the range of metabolites and were not only low values.

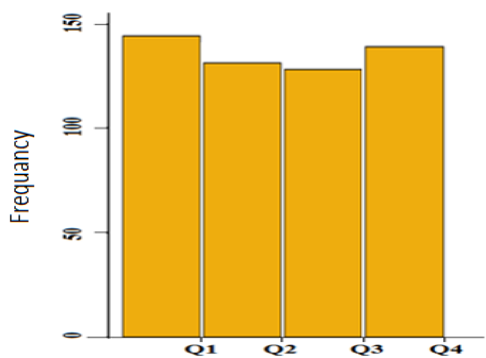

**Figure 3**. Blue: empirical distribution of observed metabolites. Red: empirical distribution of missing metabolites. The captions represent the name of metabolites and the number of missing values.

**An optimal approach for imputation:** To obtain an optimal approach for imputation in metabolomics study, we conducted a cross validation based analysis. We used 39 metabolites measured for 1977 individuals who had no missing values. While the distribution of the rate of missing values over a range of metabolites is approximately uniform, for each metabolite, we considered 14 equal intervals across the range of metabolite and randomly selected 10% of measured metabolites in each interval (14 values) and set them as missing values. To impute those missing values, we utilized five approaches that are widely applied: Iterative Robust Model-based Imputation (IR-MI) [15], which each iteration uses one variable as an outcome and the remaining variables as predictors. Multiple Imputation (MU-IM) [16], which includes multiple imputations of incomplete multivariate data values in place of missing values by running a bootstrapped EM algorithm. Maximum Likelihood estimation for multivariate normal data (ML-ES) [17], which is focused on a complete variance/covariance matrix based on maximum likelihood values imputed. K nearest neighbor (KNN) [18], which assumes data are missing at random and missing data depends on the observed data. The KNN approach is able to take advantage of multivariate relationships in the completed data. Finally, Random Forest approach [19], which is a combination of tree predictors such that each tree depends on the values of a random vector sampled independently and with the same distribution for all trees in the forest. For imputation using these approaches, we used the R packages irmi [20], amelia [21], Mvnmle [22], SeqKnn [22], and missForest [23] respectively. The performance of these methods was evaluated in terms of mean square imputed errors (MSIE) of 40 times repeated simulation scenario. Among those methods, the KNN algorithm, which uses sequential imputation, outperformed the other methods. Although it was slightly better than RF, it remarkably outperforms the other approaches. Figure 4 shows the performance of the models for K=5 as the parameter of KNN, while K in (6 to 12) did not show significant differences in our analysis.

KNN imputation based on correlation distance matrix is a simple but efficient approach that attempts to preserve original data structure and avoids distorting the distribution of imputed variables. The KNN algorithm starts from a complete subset of the data set, Xc, and sequentially estimates the missing values for an incomplete observation, x*, by minimizing the determinant of the covariance of the augmented data matrix, X* = [Xc; x']. Then the observation x* is added to the complete data matrix, and the algorithm continues with the next observation [24].

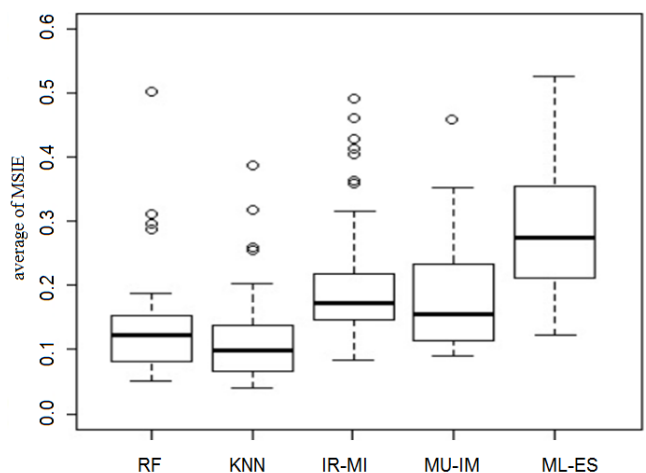

**Figure 4.** Performance of imputation methods as a function of MSIE. Y-axes: the average of MSIE over 40 sets of simulation.

As mentioned, to preserve characteristics and relationship between different metabolites, KNN takes into account the correlation among metabolites as similarity criteria. To assess how the correlation affects the imputation performance, we illustrated a simulation analysis similar to the aforementioned scenario for a set of metabolites that were selected with different correlation cutoff (0.5, 0.4, 0.3). Figure 5 demonstrates the average of correlation between imputed values and observed values of metabolites over 40 sets of simulation for each cutoff. Using this result, we imputed metabolites that showed at least 30% correlation with other metabolites and discarded the others from the analysis to avoid inaccurate imputation.

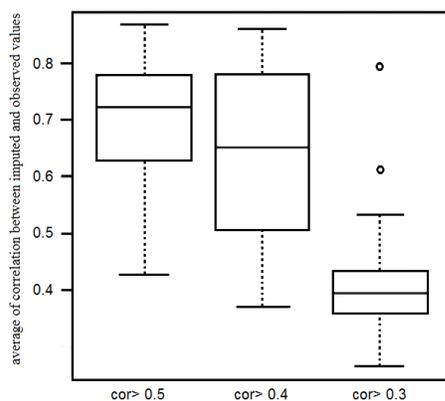

**Figure 5.** Performance of the KNN algorithm for different correlation cutoff (x-axes) among metabolites. Y-axes: the average of correlation among imputed and observed values over 40 sets of simulation.

**Discussion**

Identification of the nature of missing values and imputation are the major tasks in data pre-processing since no presentation of metabolomic data analysis is complete without careful consideration of missing data. The aim of this manuscript was to introduce novel techniques for missing value imputation by taking advantage of dozens-to-hundreds of replication samples. The introduced techniques can be generalized to a variety of studies, although we applied our techniques to metabolomics data.

Missing values widely occur in mass spectrometry metabolomics datasets due to a variety of reasons, from biology to totally technical reasons [1-3]. After the pre-processing of metabolomics data including baseline correction, noise reduction, smoothing, peak detection, and alignment [4-7], some other steps, such as imputation and transformation, are required to prepare the data for analysis. Missing value imputation has been addressed through a wide range of approaches. However, most attempts for metabolomic imputation involve easy approaches, such as using the mean, median, or lowest value of measured metabolites.

Each data set is unique, and missing values imputation needs to be carried out carefully. In this study, we introduced some statistical techniques for imputation based on the fact that some data sets include replication samples. We estimated the distribution of missing values in a set of 97 individuals with replication samples and noticed that the rate of missing values across the range of metabolites was approximately scattered as uniform. The results contradict the common belief that missing metabolite values are low values. Therefore, replacing missing data with the lowest value imposes biases to data analyses. Furthermore, based on our assessments, we observed that different transformations were required for different metabolites to be transformed to normal, and log transformation did not normalize all metabolites.

After identifying the nature of missing values in our data set, we conducted a simulation study to compare the performance of different imputation approaches and identify an optimal imputation method. To this end, we conducted a simulation study and compared five different imputation methods, IR-MI, MU-IM, ML-ES, KNN, and RF using the distribution of missing values. Through this study, we determined the performance of each approach and selected the optimal approach for imputation of metabolomics data. Among those five imputed methods, the KNN algorithm showed the best performance for metabolite imputation. After preparing the metabolomics data using the techniques introduced here, we analyzed metabolomic data with the CCRS approach (a penalization approach for rare variant selection) [25-26] and the G-DAG algorithm (a systems approach established in Mendelian randomization and Bayesian graphical modeling) [27]. The results of the systematic analysis of the metabolites imputed here are provided in applications [28-32] with clinically validated novel findings [33].

**Conclusion**

Using replication samples, we could identify the nature of missing values through finding the empirical distribution of metabolites. The result did not support the assumption that metabolites are missing because they are low. Therefore, replacing missing values with the lowest values can mislead the analysis and results in inaccurate conclusions. We observed that the rate of missing values is approximately distributed as uniform across the range of metabolites. Based on this fact, our conducted simulation study suggested the KNN algorithm as an optimal approach for metabolomics imputation.

**List of abbreviations**

Kullback-Leibler divergence (KLD)

K nearest neighbor (KNN)

Iterative Robust Model-based Imputation (IR-MI)

Multiple Imputation (MU-IM)

Maximum Likelihood estimation for multivariate normal data (ML-ES)

Mean square imputed errors (MSIE)

Random Forest (RF)

**Availability of data and materials:** Data access can be provided by dbGaP through the following link https://www.ncbi.nlm.nih.gov/projects/gap/cgi-bin/study.cgi?study_id=phs000399.v1.p2.

**Competing interests:** There is no conflict of interest.

**Funding:** Metabolomic research in ARIC was partially supported by the National Human Genome Research Institute (3U01HG004402-02S1). Azam Yazdani was supported by a training fellowship from the Keck Center for Interdisciplinary Bioscience Training of the Gulf Coast Consortia (RP140113). Authors' Contributions: The authors have contributed to the manuscript equally.

**Acknowledgements:** The authors thank Dr. Eric Boerwinkle for helpful discussions and providing an opportunity to carry out this research. The authors also thank the staff and participants of the Atherosclerosis Risk in Communities (ARIC) Study for their important contributions.

**Conflict of Interest:** Authors declare that there is no conflict of interest.